\documentclass[12pt]{article} 
\usepackage{epsfig} 

\def\bea {\begin{eqnarray}}
\def\eea {\end{eqnarray}}
\def\be{\begin{equation}}
\def\ee{\end{equation}}

\begin{document}\begin{titlepage} 

\begin{flushright}
JLAB-THY-07-721
\end{flushright} 

\begin{center}\vspace{5mm}\Large{\bf Resolution of the Proton Spin Problem}
\\
\vskip 0.5cm  
{\large F. Myhrer$^{a}$ and A.W. Thomas$^{b,c}$}
\vskip 0.5cm
$^{a}$~{\large Department of Physics and Astronomy,
University of South Carolina, 
Columbia, South Carolina 29208 }\\
$^{b}$~{\large Jefferson Lab, 12000 Jefferson Ave., \\ 
Newport News VA 23606}\\
$^{c}$~{\large College of William and Mary, Williamsburg VA 23187}
\end{center}

\centerline{(\today) }
\vskip 1cm
\begin{abstract}
A number of lines of investigation into the structure of the nucleon 
have converged to the point where we believe that one has a consistent 
explanation of the well known proton spin crisis. 
\end{abstract}
\end{titlepage}
\newpage
There is no more fundamental challenge for strong 
interaction physics than mapping 
the distribution of energy, momentum, spin and angular 
momentum onto the quarks 
and gluons that compose the nucleon. For the past two decades 
there has been a 
tremendous level of activity associated with the latter two, sparked by the 
discovery, almost 20 years ago, by the European Muon Collaboration (EMC) 
of a proton ``spin crisis''~\cite{Ashman:1987hv}. 
Much of the early theoretical effort was focused on the important task of 
understanding the role of polarized gluons and the axial anomaly in resolving 
this crisis. Impressive experimental work at CERN, DESY, JLab, RHIC and SLAC 
has established a number of important pieces of the information needed to 
understand the puzzle. 

According to EMC~\cite{Ashman:1987hv} 
the experimental indication 
was that 
the quark spin was near 
zero: $14 \pm 9 \pm 21$\%. 
This led to 
the exciting 
possibility~\cite{Efremov:1988zh,Altarelli:1988nr,Bodwin:1989nz,Bass:1991yx} 
that the proton might contain a substantial 
quantity of polarized glue which could contribute 
to reducing the quark spin through 
the famous U(1) 
axial anomaly. It has taken almost 20 years to 
investigate this fascinating possibility 
experimentally  
and there are still important 
measurements underway. 
Nevertheless, it is already 
clear that the gluon spin
is nowhere near as large as would be required 
to explain the spin problem. 
{}For example, the most recent measurements of inclusive $\pi^0$ 
jets at RHIC are best fit with 
$\Delta G = 0$~\cite{Jacobs-Pacific,Abelev:2006uq} 
and Bianchi~\cite{Bianchi,Airapetian:2007mh} 
reported $\Delta G /G \sim 0.08$ at Pacific-SPIN07.

As the accuracy of experimental 
investigation of the spin of the proton has 
increased, the fraction of the spin carried by 
quarks has moved significantly far 
towards the top of the range 
quoted by EMC. 
%
We now know that the sum of the helicities 
of the quarks in the proton corresponds to about a third its total 
spin~\cite{Kabuss_Pacific,Ageev:2005gh}, 
\begin{equation}
\Sigma_{\rm inv} = 0.33 \pm 0.03 {\rm (stat.)} \pm 0.05 {\rm (syst.)} \, ,
\label{eq:data}
\end{equation}
considerably higher than the initial EMC suggestion. 
Nevertheless, the modern 
value is still astonishingly small. 

The apparent failure of polarized glue as an explanation for the 
spin problem leads us to focus again on suggestions 
made soon after the EMC 
announcement~\cite{Myhrer:1988ap,Schreiber:1988uw,Hogaasen:1988}, 
which were based on physics that is 
more familiar to those modeling non-perturbative QCD. As we shall explain, 
these ideas have important implications for experimental efforts in this area. 
In particular, they suggest that most of the missing spin of the proton 
must be carried as orbital angular momentum by the valence quarks, which 
in turn makes the study of Generalized Parton Distributions(GPDs) after the 
12 GeV Upgrade at JLab extremely interesting.

We begin our 
discussion by summarizing the key physics leading to the observed quark 
spin, $\Sigma_{\rm inv}$, before explaining each term in more detail. 
There are three factors which, in our view, are needed in order to understand 
the data:
\begin{itemize}
\item the relativistic motion of the valence quarks
\item the virtual excitation of anti-quarks in low-lying 
p-states through the one-gluon-exchange hyperfine interaction -- 
in nuclear physics terms this would be termed an exchange current correction
\item the pion cloud of the nucleon.
\end{itemize}
These three pieces of physics, tested in many independent ways, all 
have the effect of converting quark spin to orbital angular momentum. The first 
reduces the spin by about one third, the second yields a reduction 
by an amount of order 0.15 and the third gives a multiplicative reduction by 
a factor of order 0.80 -- the details and estimates of uncertainties are given 
below. Altogether, these effects reduce the fraction of the proton spin carried 
by its quarks to about one third, in very good agreement with the data.

We now present some details of these three major reduction factors,  
which lead to the small value of $\Sigma_{\rm inv}$.  

\vspace{3mm}

1.  Relativistic Valence Quark Motion  \\ 
This effect was well understood even at the time of 
the EMC discovery. A spin-up, light quark in an s-state, moving in a confining 
potential, has a lower Dirac component in which the quark is in p-wave. 
The angular momentum coupling is such that for this component the spin is 
preferably down and reduces the ``spin content" of the valence quarks. 
In the bag model, for example, where the massless quark's ground state 
energy equals $\Omega /R \simeq 2.043/R$, the reduction factor 
$B= \Omega /3 (\Omega -1) \simeq 0.65$.  
The same factor reduces the value of $g_A$ from 5/3 to $\simeq$ 1.09 in a
bag model and this value changes little if one uses 
typical light quark current masses. 
Even in more modern relativistic models, where quark confinement is 
simulated by forbidding on-shell propagation through proper-time 
regularization, the reduction factor is very similar -- e.g., in 
Ref.~\cite{Cloet:2007em} $\Delta u + \Delta d$ is 0.67. 
In terms of following where the nucleon spin has gone, the relativistic 
motion transfers roughly 35\% of the nucleon spin from quark spin 
to valence quark orbital angular momentum.

\vspace{3mm}

\begin{figure}
\begin{center}
\epsfig{file=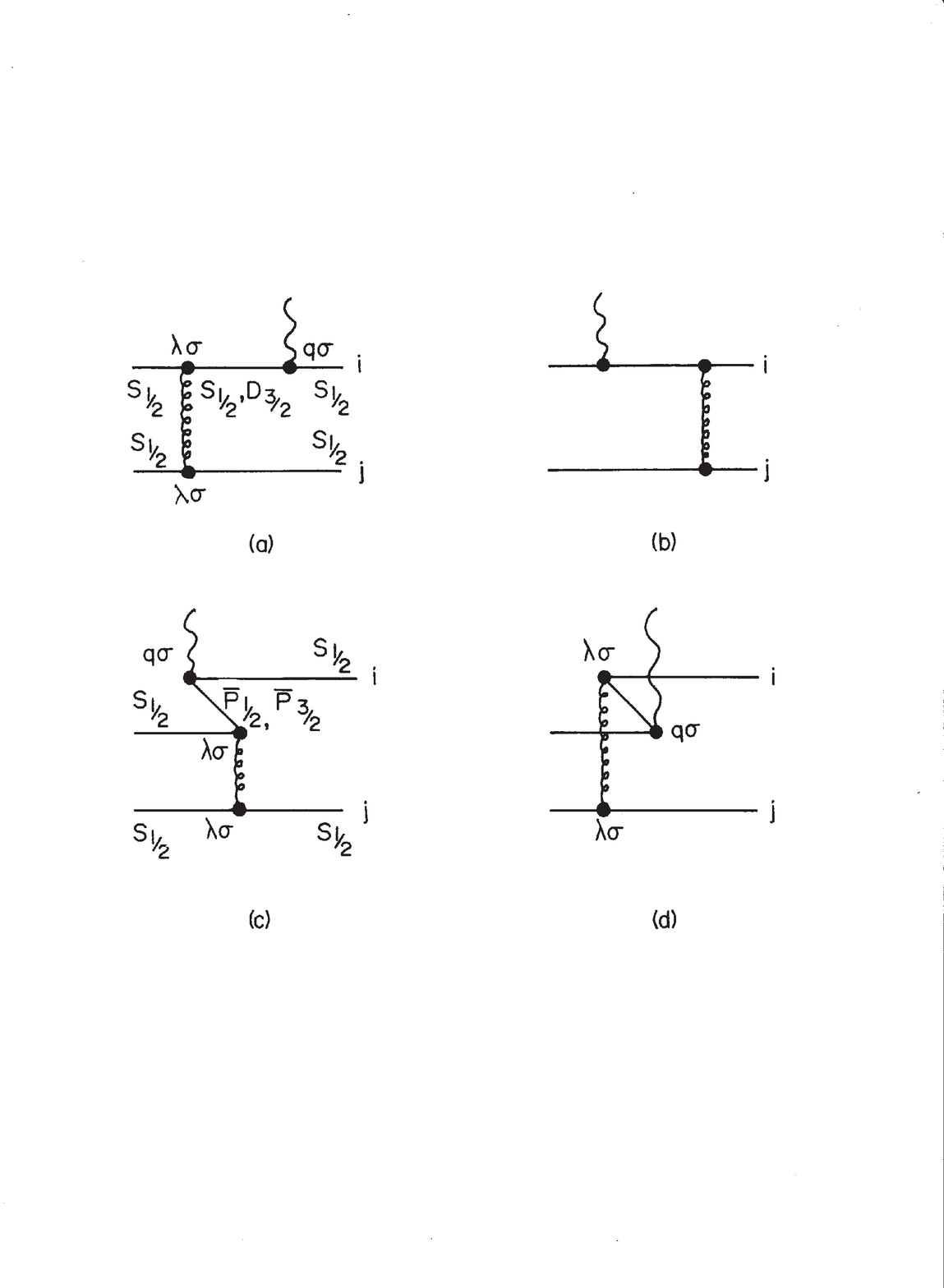, width=6cm 
}
\end{center}
\caption[]{One gluon exchange graphs contributing to the proton spin. 
}
\label{fig:I}
\end{figure}

2. The One-Gluon-Exchange Hyperfine Interaction \\ 
It is well established that the spin-spin interaction between quarks in a baryon, arising from 
the exchange of a single gluon, 
explains a major part of the mass 
difference between the octet 
and decuplet baryons -- e.g., the nucleon-$\Delta$ 
mass difference~\cite{De Rujula:1975ge,Chodos:1974pn}. This
spin-spin interaction must therefore also play a 
role when an external probe interacts with 
the three-quark baryon state. That is,   
the probe not only senses a single quark 
current but a two-quark current 
as well. 
The latter has an intermediate quark propagator connecting 
the probe and the spin-spin interaction vertices, 
and is similar to 
the exchange-current corrections which are 
well known in nuclear physics. 
In the 
context of spin sum rules 
the probe couples to the 
various axial currents in the nucleon. 
In the  case of the two-quark current, 
first investigated in detail in 
Ref.~\cite{Hogaasen:1988jd}, using the MIT bag model, 
the quark propagator was written as a sum over 
quark eigenmodes and the dominant 
contributions were found to come from the 
intermediate p-wave anti-quark states. 
The primary focus of Ref.~\cite{Hogaasen:1988jd} was 
actually the one-gluon-exchange corrections 
to the magnetic moments and semi-leptonic 
decays of the baryon octet. For example, this 
exchange current correction is vital to 
understand the unusual strength of the decay 
$\Sigma^- \rightarrow n + e^- + \bar{\nu}_e$. 

Myhrer and Thomas~\cite{Myhrer:1988ap} 
realized the importance 
of this correction to the flavor singlet axial 
charge and hence to the proton 
spin, finding that it reduced the fraction 
of the spin of the nucleon carried by quarks,  
calculated in the naive bag model by 0.15,  
i.e., $\Sigma \rightarrow \Sigma - 3G$~~\cite{Myhrer:1988ap}.
The correction term,  
$G$, is proportional to 
$\alpha_s$ times certain bag model matrix 
elements~\cite{Hogaasen:1988jd}, where $\alpha_s$ is 
determined by the ``bare" nucleon-$\Delta$ mass difference.
Again, the spin lost by the quarks is compensated 
by orbital angular momentum of the  
quarks and anti-quarks (predominantly $\bar{u}$ 
in the p-wave).

\vspace{3mm}
3. The pion cloud \\   
We know that many static baryon observables, 
such as the baryon magnetic moments and charge 
distributions, acquire important contributions from their  
pion cloud~\cite{Leinweber:2001ui}. 
This pion cloud is an effective description of 
the quark-antiquark excitations 
which are required by the chiral symmetry of QCD. 
In fact, describing a physical 
nucleon as having a pion cloud which interacts with the valence quarks of 
the quark core (the  ``bare'' nucleon), 
in a manner dictated by the 
requirements of chiral symmetry, has been very successful 
in describing 
the properties of the nucleon~\cite{Theberge:1980ye,Thomas:1982kv,mbx:1981}. 
The cloudy bag model (CBM)~\cite{Theberge:1980ye,Thomas:1982kv} 
reflects this description 
of the nucleon and in this model the  nucleon consists of a bare nucleon, 
$| N>$, with a probability $Z \sim 1-P_{N\pi} -P_{\Delta \pi} \, \sim 
\, 0.7$, in addition to being described 
as a nucleon ($N$) and a pion and a $\Delta$ and a pion, with probabilities 
$P_{N\pi} \sim 0.20-0.25 $ and $P_{\Delta \pi} \sim 0.05-0.10 $, respectively. 
The phenomenological constraints on these probabilities were 
discussed, for example, in Refs.~\cite{Speth:1996pz,Melnitchouk:1998rv}. 
One of the most famous 
of these constraints is associated 
with the excess of $\bar{d}$ over $\bar{u}$ 
quarks in the proton, predicted 
on the basis of the CBM~\cite{Thomas:1983fh}. 
Indeed, to first order the integral 
of $\bar{d}(x) - \bar{u}(x)$ 
is 2/3 $P_{N\pi}$, which is 
experimentally consistent with the 
range just quoted~\cite{Arneodo:1996kd}.

The pion cloud effect was investigated 
early by Schreiber and Thomas, who wrote the corrections to the 
spin sum-rules for the proton and 
neutron explicitly in terms of the 
probabilities set out above~\cite{Schreiber:1988uw}. {}For our purposes 
it is helpful to summarize the results of Ref.~\cite{Schreiber:1988uw} for 
the proton and neutron. 
The pion cloud correction to the 
flavor singlet combination modifies the 
proton spin in the following manner:
\begin{equation}
\Sigma \rightarrow  \left(Z - 
\frac{1}{3} P_{N \pi} +\frac{5}{3} P_{\Delta \pi} \right) \Sigma \, .
\label{eq:pion}
\end{equation}
{}From the point of view of 
the spin problem, the critical 
feature of the pion cloud is 
that the 
coupling the spin of the nucleon 
and the orbital angular 
momentum of the pion in the 
$N \pi$ Fock state favors a spin 
down nucleon and a pion with +1 unit of orbital angular momentum. 
This too has the effect of replacing 
quark spin by quark and 
anti-quark orbital angular 
momentum. 
Note that in the $\Delta \pi$ Fock component 
the spin of the baryon tends to 
point up (and the pion angular momentum down), 
thus enhancing the quark spin. 
Nevertheless, the wave function 
renormalization factor, $Z$, dominates, 
yielding a reduction by a factor between 0.7 and 
0.8 for the range of probabilities quoted above.  

\vspace{3mm} 
4. Discussion \\
%
The corrections described here, 
which arise from either the pion cloud or 
gluon exchange, lead to a significant movement from 
the theoretically expected value of $\Sigma \simeq$ 
0.65 (because of relativistic motion of the quarks) towards the experimental value. The 
one-gluon-exchange correction moves $\Sigma$
down to 0.50, while the pion cloud 
correction reduces $\Sigma$ to between 0.46 and 0.52. 
At the time these corrections were first discussed, 
neither the 
one-gluon-exchange correction, nor the pion 
cloud, seemed to yield a correction large   
enough to be relevant to resolving the crisis.
Furthermore, we  
were reticent to combine 
the one-gluon-exchange and pion cloud 
corrections as it was expected that the latter 
might contribute a substantial fraction of the 
observed splitting between the N and $\Delta$, 
which would in turn reduce the strength of 
the one-gluon-exchange term. However, progress 
in the analysis of lattice QCD calculations, 
especially in the last few years, 
changes the situation. 
In particular, the chiral analysis of 
quenched and full QCD data for the N and $\Delta$
masses as a function of quark 
mass~\cite{Young:2002cj,Young:2002rx}, 
has led to the conclusion that pion 
effects likely contribute 
50 MeV or less of the observed 
300 MeV mass difference. As a result we no longer 
need to worry about significant double counting 
and can therefore 
combine the one-gluon-exchange and pion cloud 
corrections to the quark spin sum. 

In fact, 
it is apparent that if we combine 
the one-gluon-exchange and pion cloud 
corrections, which we have just summarized,  
one finds a value for $\Sigma$ 
between 0.35 ($P_{N \pi} =0.25, 
P_{\Delta \pi}=0.05$) and 0.40 ($P_{N \pi} =0.20,
P_{\Delta \pi}=0.10$) 
in excellent agreement with the modern data.

Until now we have used the symbol $\Sigma$ to 
denote the quark spin evaluated within a model. 
This has variously been identified in the 
literature with either $\Sigma_{\rm inv}$~\cite{Hogaasen:1995nn} or 
with the value measured at some relatively 
low scale, $\mu^2 < 1$ GeV$^2$. The difficulty 
arises because the flavor singlet 
spin operator has a divergence 
in QCD involving an operator of dimension 4, 
so that its matrix element is not renormalization 
group invariant. Fortunately, at 
the current level of precision, this issue is of 
little practical importance for us. In the former case, 
the model result is 
$\Sigma \in (0.35,0.40)$, which agrees 
very well with the experimental value 
$\Sigma_{\rm inv}$ -- c.f. Eq.(\ref{eq:data}). 
In the latter case, 
which is motivated by the observation that a valence 
dominated quark model can only match experiment 
for parton distribution functions at a relatively 
low scale~\cite{Parisi:1973nx,Signal:1988vf,Gluck:1988ey}, 
the calculated value of the quark spin would 
need to be multiplied by a non-perturbative 
factor involving the QCD $\beta$--function and the 
anomalous dimension, $\gamma$, of the flavor singlet axial charge, 
which has been calculated to three loops by 
Larin and 
Vermaseren~\cite{Larin:1991tj,Larin:1994va}. As 
this factor is truly non-perturbative, its 
evaluation through even three-loop perturbation 
theory is at best 
semi-quantitative~\cite{Bass:2002mv}. Nevertheless, 
it is rigorously less than unity and an evaluation 
at three-loops gave a value of order 0.6--0.8~\cite{Hogaasen:1995nn}. 
Multiplying the quark spin 
obtained above by this factor yields a value for
$\Sigma
\in (0.21,0.32)$, which is also 
in excellent agreement with the current experimental value.

In conclusion, the tremendous experimental progress 
aimed at resolving the spin problem has 
established that the quarks carry about one third 
of the spin of the nucleon and that the 
polarization of the gluons is most likely 
too small to 
account for the difference. Instead, well known 
aspects of hadron structure involving its pion cloud 
and the hyperfine interaction mediated by 
one-gluon exchange, in combination with the 
relativistic motion of the confined quarks, are 
able to explain the modern data very satisfactorily. 
As a consequence of these new insights, we expect 
that the missing spin should be accounted for by 
the orbital angular momentum of the quarks and 
anti-quarks -- the latter associated with the 
pion cloud of the nucleon and the p-wave anti-quarks 
excited by the one-gluon-exchange 
hyperfine interaction. Finally, we note 
that the exploration of the 
angular momentum carried by the quarks 
and anti-quarks is a major aim of the 
scientific program associated with the 12 GeV 
Upgrade at Jefferson Lab. 

{\bf Acknowledgements}

This work was supported by the US Department 
of Energy, Office of Nuclear Physics, through 
contract no. DE-AC05-06OR23177, under which 
Jefferson Science Associates operates Jefferson 
Lab. and by the NSF grant PHY-0457014.


\vspace{3mm} 
\begin{thebibliography}{99} 
%
\bibitem{Ashman:1987hv} J.~Ashman {\it et al.}  [European Muon Collaboration],
Phys.\ Lett.\  B {\bf 206}, 364 (1988).
%
\bibitem{Efremov:1988zh}
A.~V.~Efremov and O.~V.~Teryaev,
Report JINR-E2-88-287.
%
\bibitem{Altarelli:1988nr}
  G.~Altarelli and G.~G.~Ross,
  Phys.\ Lett.\  B {\bf 212}, 391 (1988).
%
\bibitem{Bodwin:1989nz}
G.~T.~Bodwin and J.~W.~Qiu,
Phys.\ Rev.\  D {\bf 41}, 2755 (1990).
%
\bibitem{Bass:1991yx}
S.~D.~Bass, B.~L.~Ioffe, N.~N.~Nikolaev and A.~W.~Thomas,
J.\ Moscow.\ Phys.\ Soc.\  {\bf 1}, 317 (1991).
%
\bibitem{Jacobs-Pacific}
W.~Jacobs, invited talk presented at 
Pacific-SPIN07, July 30-August 2, 2007 
({\it op.cit.}).
%
\bibitem{Abelev:2006uq}
B.~I.~Abelev {\it et al.}  [STAR Collaboration],
Phys.\ Rev.\ Lett.\  {\bf 97}, 252001 (2006)
[arXiv:hep-ex/0608030].
%
\bibitem{Bianchi}
N. Bianchi, for the Hermes Collaboration,
invited talk presented at 
Pacific-SPIN07, July 30-August 2, 2007 
({\it op.cit.}).
%
\bibitem{Airapetian:2007mh}
A.~Airapetian {\it et al.}  [HERMES Collaboration],
``Precise determination of the spin structure 
Phys.\ Rev.\  D {\bf 75}, 012007 (2007).
%
\bibitem{Kabuss_Pacific}
E.-M.~Kabuss for the COMPASS Collaboration, invited 
talk presented at Pacific-SPIN07, Vancouver, July 30-August 2, 2007:
http://www.triumf.info/hosted/pacspin07/program.htm
%
\bibitem{Ageev:2005gh}
E.~S.~Ageev {\it et al.}  [COMPASS Collaboration],
Phys.\ Lett.\  B {\bf 612}, 154 (2005)
[arXiv:hep-ex/0501073].
%
\bibitem{Myhrer:1988ap}
  F.~Myhrer and A.~W.~Thomas,
  Phys.\ Rev.\  D {\bf 38}, 1633 (1988).
%
\bibitem{Schreiber:1988uw}
  A.~W.~Schreiber and A.~W.~Thomas,
  Phys.\ Lett.\  B {\bf 215}, 141 (1988).
%
\bibitem{Hogaasen:1988}
H.~H{\o}gaasen and F.~Myhrer, 
  Phys.\ Lett.\  B {\bf 214}, 123 (1988)..
%
\bibitem{Cloet:2007em}
  I.~C.~Cloet, W.~Bentz and A.~W.~Thomas,
  arXiv:0708.3246 [hep-ph].
%
\bibitem{De Rujula:1975ge}
  A.~De Rujula, H.~Georgi and S.~L.~Glashow,
  Phys.\ Rev.\  D {\bf 12}, 147 (1975).
%
\bibitem{Chodos:1974pn}
  A.~Chodos, R.~L.~Jaffe, K.~Johnson and C.~B.~Thorn,
  Phys.\ Rev.\  D {\bf 10}, 2599 (1974).
%
\bibitem{Hogaasen:1988jd}
  H.~H{\o}gaasen and F.~Myhrer,
Phys.\ Rev.\  D {\bf 37}, 1950 (1988). 
%
\bibitem{Leinweber:2001ui}
  D.~B.~Leinweber, A.~W.~Thomas and R.~D.~Young,
  Phys.\ Rev.\ Lett.\  {\bf 86}, 5011 (2001)
  [arXiv:hep-ph/0101211].
%
\bibitem{Theberge:1980ye}
  S.~Theberge, A.~W.~Thomas and G.~A.~Miller,
  Phys.\ Rev.\  D {\bf 22}, 2838 (1980)
  [Erratum-ibid.\  D {\bf 23}, 2106 (1981)].
%
\bibitem{Thomas:1982kv}
  A.~W.~Thomas,
  Adv.\ Nucl.\ Phys.\  {\bf 13}, 1 (1984).

\bibitem{mbx:1981} F. Myhrer, G.E. Brown and Z. Xu, 
Nucl. Phys. A {\bf 362}, 317 (1981). 

%
\bibitem{Speth:1996pz}
  J.~Speth and A.~W.~Thomas,
  Adv.\ Nucl.\ Phys.\  {\bf 24}, 83 (1997).
%
\bibitem{Melnitchouk:1998rv}
  W.~Melnitchouk, J.~Speth and A.~W.~Thomas,
  Phys.\ Rev.\  D {\bf 59}, 014033 (1999)
  [arXiv:hep-ph/9806255].
%
\bibitem{Thomas:1983fh}
  A.~W.~Thomas,
  Phys.\ Lett.\  B {\bf 126}, 97 (1983).
%
\bibitem{Arneodo:1996kd}
  M.~Arneodo {\it et al.}  [New Muon Collaboration],
  Nucl.\ Phys.\  B {\bf 487}, 3 (1997)
  [arXiv:hep-ex/9611022].
%
%
\bibitem{Young:2002cj}
  R.~D.~Young, D.~B.~Leinweber, A.~W.~Thomas and S.~V.~Wright,
  Phys.\ Rev.\  D {\bf 66}, 094507 (2002)
  [arXiv:hep-lat/0205017].
%
\bibitem{Young:2002rx}
  R.~D.~Young, D.~B.~Leinweber, A.~W.~Thomas and S.~V.~Wright,
  Nucl.\ Phys.\ Proc.\ Suppl.\  {\bf 109A}, 55 (2002).
%
\bibitem{Parisi:1973nx}
G.~Parisi,
Phys.\ Lett.\  B {\bf 43}, 207 (1973).
%
\bibitem{Signal:1988vf}
A.~I.~Signal and A.~W.~Thomas,
Phys.\ Lett.\  B {\bf 211}, 481 (1988).
%
\bibitem{Gluck:1988ey}
M.~Gluck and E.~Reya,
Z.\ Phys.\  C {\bf 43}, 679 (1989).
%
\bibitem{Larin:1991tj}
S.~A.~Larin and J.~A.~M.~Vermaseren,
Phys.\ Lett.\  B {\bf 259}, 345 (1991).
%
\bibitem{Larin:1994va}
S.~A.~Larin, 
%
%
Phys.\ Lett.\ B {\bf 334}, 192 (1994). 
%
\bibitem{Bass:2002mv}
S.~D.~Bass, R.~J.~Crewther, F.~M.~Steffens and A.~W.~Thomas,
Phys.\ Rev.\  D {\bf 66}, 031901 (2002)
[arXiv:hep-ph/0207071].
%
\bibitem{Hogaasen:1995nn}
H.~H{\o}gaasen and F.~Myhrer,
Z.\ Phys.\  C {\bf 68}, 625 (1995)
[arXiv:hep-ph/9501414].
%
\end{thebibliography}
\end{document}